\documentclass[]{elsart}

\usepackage{graphicx}
\usepackage{subfigure}
\usepackage[latin1]{inputenc}
\usepackage{amsmath}
\usepackage{cite}
\usepackage{xspace}
\usepackage{amssymb}

\newcommand{\ie}{{i.e.}\@\xspace}
\newcommand{\cf}{{cf.}\@\xspace}

\newcommand{\nX}{n_{\mathrm{Sc}}}
\newcommand{\nSc}{n_{\mathrm{Sc}}}

\begin{document}
\begin{frontmatter}
	\title{Using cluster dynamics to model electrical resistivity
	measurements in precipitating Al-Sc alloys}
	\author{Emmanuel Clouet\corauthref{email}} and
	\author{Alain Barbu}
	\corauth[email]{emmanuel.clouet@cea.fr}
	\address{Service de Recherches de Métallurgie Physique, CEA/Saclay, \\
	91191 Gif-sur-Yvette, France}
	\begin{abstract}
		Electrical resistivity evolution during precipitation in Al-Sc alloys 
		is modeled using cluster dynamics. This mesoscopic modeling 
		has already been shown to correctly predict the time evolution 
		of the precipitate size distribution.
		In this work, we show that it leads too
		to resistivity predictions in quantitative agreement with experimental data.
		We only assume that all clusters contribute
		to the resistivity and that each cluster contribution is proportional 
		to its area.
		One interesting result is that the resistivity excess 
		observed during coarsening mainly arises from large clusters
		and not really from the solid solution.
		As a consequence, one cannot assume that resistivity asymptotic behavior
		obeys a simple power law as predicted by LSW theory for the solid 
		solution supersaturation.
		This forbids any derivation of the precipitate interface free energy
		or of the solute diffusion coefficient from resistivity experimental data 
		in a phase-separating system like Al-Sc supersaturated alloys.
	\end{abstract}
\end{frontmatter}

\section{Introduction}

Resistivity measurements are usually a convenient way 
to follow precipitation kinetics in phase-separating systems.
Indeed, assuming that resistivity is proportional to the solid solution
solute content as it is usually done\cite{ROS87}, 
one gets direct access to the solute supersaturation.
The so-called LSW theory (Lifshitz and Slyozov\cite{LIF61} and Wagner\cite{WAG61})
as extended by Ardell \cite{ARD67} allows then to deduce from these measurements
key parameters like the solubility, the precipitate interface free energy 
and the solute diffusion coefficient.

Numerous such measurements exist in Al-Sc alloys 
\cite{DRI84,JO93,NAK97,ZAK97,WAT04,ROY05c}.
They have been intended to characterize the precipitation of the
Al$_3$Sc L1$_2$ structure in aluminum alloys.
In a previous study\cite{CLO05}, we used cluster dynamics 
to model precipitation in Al-Zr and Al-Sc alloys.
This mesoscopic modeling was shown to provide predictions
of the precipitate size distributions in quantitative agreement
with available experimental data\cite{NOV01,MAR01,MAR02}. 
It is therefore worth seeing if such an agreement can be obtained also
with resistivity measurements.
The purpose is mainly to see how predictive cluster dynamics can be, 
\ie if one can obtain reliable information concerning the precipitates 
as well as the solid solution in the whole time range of precipitation kinetics
and not only in the asymptotic limit of the coarsening stage like LSW theory does.
Confronting cluster dynamics predictions with resistivity measurements will 
validate too the multiscale approach developed for
precipitation in Al-Zr-Sc alloys \cite{CLO04,CLO04T,CLO05,CLO05b,CLO06}.
The use of resistivity experiments to validate a multiscale kinetic modeling 
has already proved its efficiency for irradiated iron \cite{FU05}.

The first part of this article explains how resistivity can be modeled 
within the framework of  cluster dynamics.
A thorough comparison between the obtained model predictions and 
available experimental data is then performed.
This allows us to conclude on the validity of the interface free energies
which are obtained through such measurements.


\section{Cluster gas model of electrical resistivity}

Cluster dynamics models the phase-separating alloy as a gas of solute clusters
which exchange solute atoms by single atom diffusion.
Clusters are assumed to be spherical and are described by a single parameter, 
the number $\nSc$ of solute atoms they contain.
The time evolution of the cluster size distribution is governed 
by a master equation which input parameters are 
the solute diffusion coefficient and the precipitate interface free energy.
For Al-Sc alloys, these needed parameters 
were directly deduced from an atomic diffusion model\cite{CLO04,CLO04T,CLO06} 
previously built for Al-Zr-Sc alloys.
It should be stressed that, although none of these input parameters were intentionally fitted,
cluster dynamics managed to reproduce reasonably experimental data
on the time evolution of the precipitate size distribution.
The reader is referred to Ref.~\citen{CLO05} for a full description of 
cluster dynamics modeling and its application to Al-Sc alloys.
We will only describe in the following how this technique can be adapted 
to model electrical resistivity of a phase-separating Al-Sc alloy.

\subsection{Cluster contribution to electrical resistivity}

So as to use cluster dynamics to simulate the evolution of the electrical resistivity
during the annealing of a supersaturated solid solution, 
we assume that the total resistivity of the cluster gas is given 
by the sum of each cluster resistivity. Doing so, we neglect any interference
between clusters. 
Therefore, the input parameters of the modeling are the contributions of each cluster,
\ie $\rho_{\nX}$ for a cluster containing $\nX$ solute atoms. 
Some calculations of this contribution exist in the literature (for a review, see Ref.~\citen{ROS87}).
Most of them have been aimed to explain the increase of resistivity 
at the beginning of the precipitation kinetics in alloys like Al(Zn), Al(Ag), Al(Cu) or Al(Cr),
showing that this resistivity increase is associated with the apparition of small clusters ($\sim10-20$~\AA).

All these calculations consider the elastic scattering of valence electrons by 
the perturbing potential due to the presence of the solute atoms composing the cluster.
They show that solute atoms which are agglomerated in a small cluster
can have a higher resistivity than the same isolated solute atoms
because of Bragg scattering.
Indeed, the smallest the cluster, the most relaxed the Bragg conditions
for the electron scattering by the perturbing potential. 
This can lead to a high resistivity for small clusters.
The scattering anisotropy increases with the cluster size and, 
as the cluster size gets of the order of the electron mean free path,
the scattering becomes Bragg-like leading 
to a small contribution of the cluster to the resistivity.


In Al-Sc alloys, such an increase of the resistivity in the beginning of the precipitation
kinetics has never been observed experimentally \cite{DRI84,JO93,NAK97,ZAK97,WAT04,ROY05c}. 
Actually, it does not seem to exist in alloys where the solute is a transition element 
and the solvent a free electron like metal\cite{MER80a}.
Therefore, precise calculations of each cluster contribution to the electrical resistivity
may not be required for Al-Sc alloys.
Moreover, these calculations are limited to small clusters as one assumes
that the perturbing potential of the cluster does not modify the Fermi surface of the solvent 
and that electrons are scattered only once by the cluster (Born approximation). 
Such assumptions are reasonable only as long as the cluster does not contain more than 
a few solute atoms ($\sim$10) but do not really apply for larger clusters.
On the other hand, for clusters with a stoichiometric composition, 
one can consider that the conductivity is infinite inside the cluster 
and that electrons are only scattered by the interface between the cluster and the matrix.
For a sharp interface, this leads to a cluster contribution to the resistivity proportional 
to its cross section\cite{ROS87} and therefore to its interface area as clusters
are assumed spherical.
This leads then to  $\rho_{\nX}=\rho_1{\nX}^{2/3}$, and
the electrical resistivity of the phase separating system at time $t$ is simply given by
\begin{equation}
	\label{eq:resistivity}
	\rho(t) = \rho^0_{\mathrm{Al}} + \rho_1 \sum_{\nX=1}^{\nX^{*}}{C_{\nX}(t) {\nX}^{2/3}}
\end{equation}
where $C_{\nX}(t)$ is the atomic fraction of clusters containing $\nX$ solute atoms.
The time evolution of this cluster size distribution 
is governed by the cluster dynamics master equations (Eq.~2 and 3 in Ref.~\citen{CLO05}).
In Eq.~\ref{eq:resistivity}, we have assumed that only clusters smaller than a critical size $\nX^{*}$
contribute to the electrical resistivity. We will see below which value this critical
size has to be given in order to fit experimental data.
$\rho^0_{\mathrm{Al}}$ is the electrical resistivity of the solvent (Al in this case).
It is temperature dependent and is known experimentally.
The only remaining unknown parameter in Eq.~\ref{eq:resistivity} is the contribution $\rho_1$ 
of a solute monomer to the electrical resistivity.

The increase of resistivity with Sc content has been measured at 77~K
by Fujikawa \etal \cite{FUJ79,JO93} who give 
$\delta\rho_{\mathrm{Sc}}=3400$~n$\Omega$m per Sc atomic fraction.
These measurements have been performed in under-saturated and therefore 
very dilute Al-Sc solid solutions.
One can reasonably assumes that most of the solute are monomers 
and ignore larger clusters\footnote{The maximal solubility limit of Sc in aluminum 
is 0.288~at.\%. We have checked with our cluster gas model of electrical resistivity 
in Al-Sc that only the monomer contribution is relevant for such a low nominal concentration.},
with the consequence that $\rho_1=\delta\rho_{\mathrm{Sc}}=3400$~n$\Omega$m.
Assuming that Matthienssen rule\cite{ROS87} is obeyed, this quantity 
does not depend on the temperature.
Different measured values of $\delta\rho_{\mathrm{Sc}}$ ranging from 3000 to 8200~n$\Omega$m
per Sc atomic fraction
can be found in the literature (for a review, see Tab.~2 in Ref.~\citen{ROY05c}). 
Nevertheless we find that the value measured by Fujikawa \etal \cite{FUJ79,JO93}
is the one that leads to the best agreement between the simulated and the experimental
resistivity variations during precipitation kinetics.

\subsection{Critical size}

\begin{figure}[!bp]
	\begin{center}
		\includegraphics[width=0.8\linewidth]{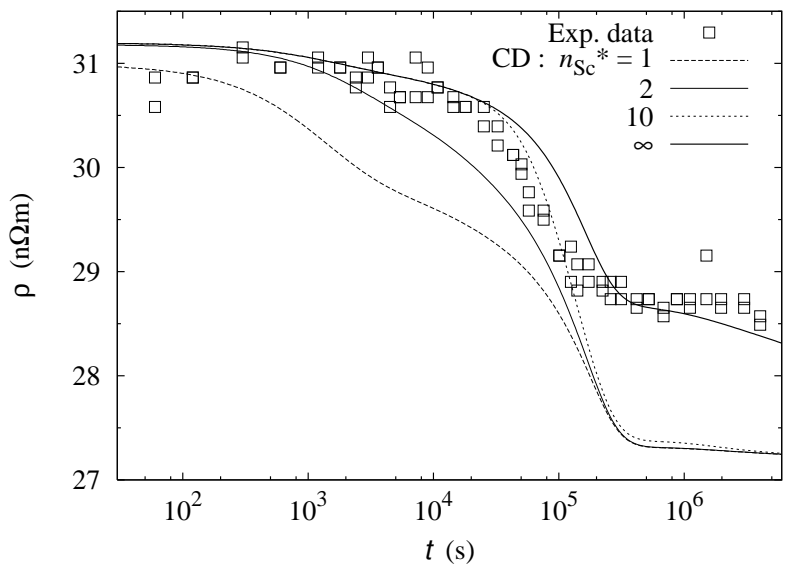}
	\end{center}
	\caption{Time evolution of the resistivity $\rho$ experimentally observed
	\cite{ROY05c} and deduced from cluster dynamics simulations for a solid solution of composition
	$x^0_{\mathrm{Sc}}=0.12$~at.\% annealed at 230°C. Different critical size $\nX^*$
	have been used in cluster dynamics simulations (Eq.~\ref{eq:resistivity}).}
	\label{fig:resistivity_nStar}
\end{figure}

\begin{figure}[!bp]
	\begin{center}
		\includegraphics[width=0.8\linewidth]{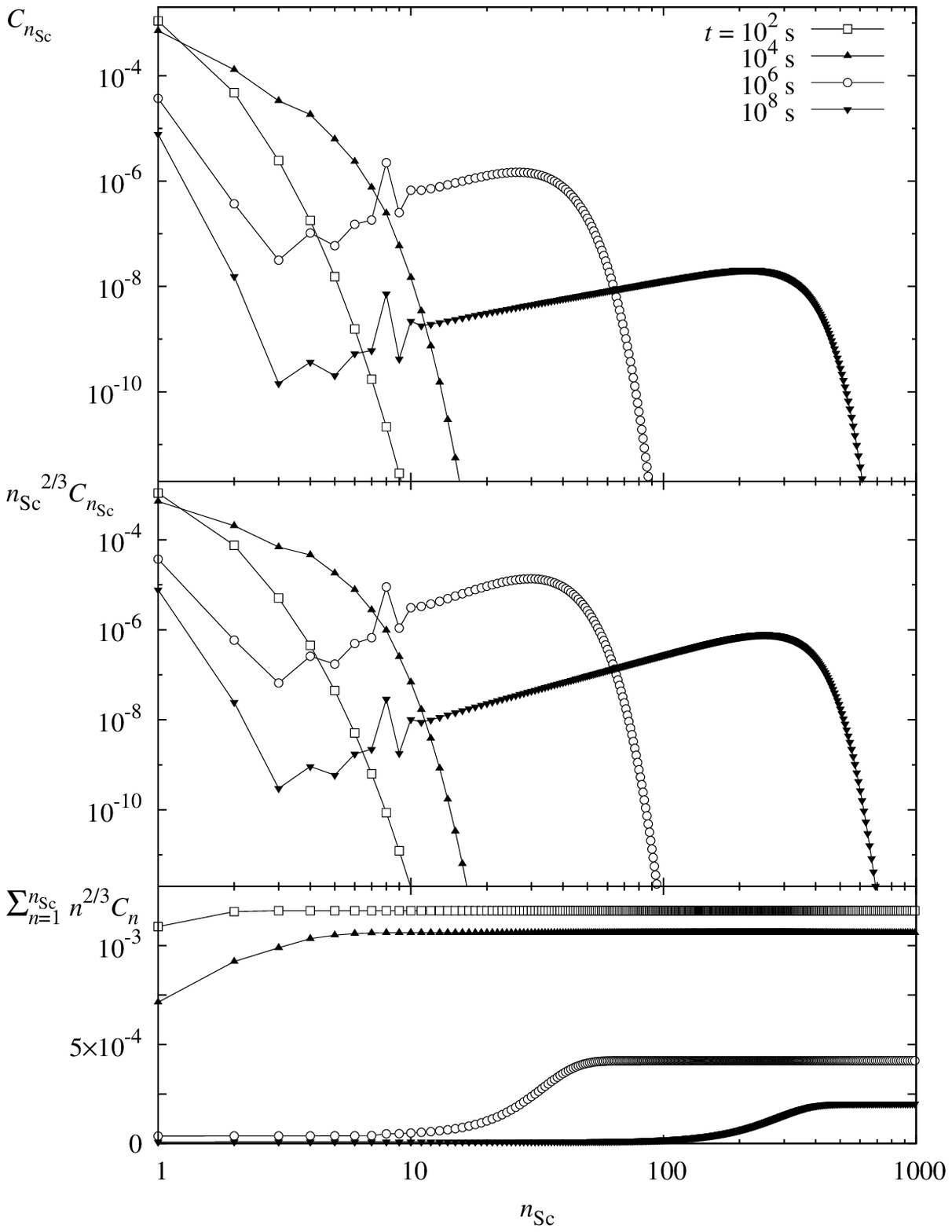}
	\end{center}
	\caption{Time evolution of the cluster size distribution ($C_{\nX}$), 
	of the cluster contribution to electrical resistivity (${\nX}^{2/3} C_{\nX}$)
	and of the cumulative contribution ($\sum_{n=1}^{\nX}{C_{n} {n}^{2/3}}$)
	in a solid solution of composition
	$x^0_{\mathrm{Sc}}=0.12$~at.\% annealed at 230°C.
	Resistivities have been normalized by $\rho_1$.}
	\label{fig:distribution}
\end{figure}

As already recalled above, the input parameters of cluster dynamics, 
\ie the precipitate interface free energy and the Sc diffusion coefficients,
were directly deduced from an atomic diffusion model \cite{CLO05}. 
The only left parameter needed to calculate electrical resistivity 
is the critical size $\nX^{*}$ appearing in Eq.~\ref{eq:resistivity}.
So as to determine this boundary between clusters which are contributing or not 
to the electrical resistivity, we compare our predictions using different values for $\nX^{*}$
with experimental data (Fig.~\ref{fig:resistivity_nStar}).
If one assumes that only monomers are contributing to the electrical resistivity ($\nX^{*}=1$),
cluster dynamics do not reproduce experimental data.
Indeed the simulated resistivity decreases too quickly during the precipitation kinetics. 
So as to obtain the plateau experimentally observed at the beginning, 
one has to take into account the contributions of small clusters ($\nX\leq10$), 
especially the dimer one.
With these contributions, the simulated resistivity decreases at the right time, 
but the final plateau experimentally observed is not reproduced. 
Indeed, cluster dynamics predicts that resistivity decreases to reach a value close 
to the one corresponding to the equilibrium solid solution, whereas experimental data
show an excess resistivity.
The final plateau corresponding to this excess resistivity can only be obtained if one
considers that all clusters are contributing to the resistivity 
($\nX^{*}=\infty$\footnote{Cluster dynamics equations can be solved only for a finite number of classes. 
The infinite size in the simulations corresponds to a maximal size which can be as large
as $\sim10^{12}$ solute atoms. We check that the concentration of clusters having this maximal size
does not evolve during the simulation. If this is not the case, the maximal size is increased.}).
With this value of the critical size, the experimental time evolution of the resistivity 
is reasonably reproduced by cluster dynamics.

So as to better understand the evolution of the electrical resistivity 
we monitor the cluster size distribution\footnote{Looking 
at the cluster size distribution, one should notice the abnormally high concentration 
of clusters containing 8 solute atoms. This is due to the low interface free energy 
associated with this size because of the existence of a compact cluster corresponding
to a cube for this size\cite{CLO04}.}
and each cluster contribution to resistivity 
(Fig.~\ref{fig:distribution}) at different times of the precipitation kinetics
for the supersaturation and the annealing temperature corresponding to Fig.~\ref{fig:resistivity_nStar}.
At the beginning of the kinetics, ($t=10^2$ or $10^4$~s), only small clusters ($\nSc\leq10$)
are present. Therefore, one can considers that resistivity only arises from these clusters.
At the very beginning ($t\leq10^2$~s), one can even only take into account mono- and dimers.
As precipitation goes on, the solid solution becomes depleted and thus the atomic fractions 
of small clusters decrease. For $t\geq10^6$~s, one can neglect the contributions 
of these small clusters to resistivity.
On the contrary, as the precipitating phase appears, more and more large clusters are present
in the system and only these large clusters contribute to the resistivity.
For instance, for $t=10^6$~s, almost all the resistivity is due to clusters
containing between 10 and 50 solute atoms. 
Thus large clusters are responsible for the excess resistivity 
observed at the end of the precipitation kinetics.
As we are in the coarsening stage by this time, the concentrations of these larges clusters
evolves very slowly, leading to a nearly stable resistivity.

\section{Comparison with experimental data}

\begin{table}[!hbtp]
	\centering
	\caption{Summary of the experimental conditions (temperature, 
		concentration $x^0_{\mathrm{Sc}}$,
		supersaturation $x^0_{\mathrm{Sc}}/x^{\mathrm{eq}}_{\mathrm{Sc}}$)
		corresponding to the available data used for the comparison
		with cluster dynamics simulations of the electrical resistivity variations
		in Al-Sc alloys.}
	\label{tab:experimental_data}
	\begin{tabular}{lccc}
		\hline
		Reference			& Temperature	& Concentration	& Supersaturation \\
		\hline
		R{\o}yset and Ryum \cite{ROY05c}	& 190°C		& 0.12~at.\%	& 3530	\\
						& 230°C		& 0.12~at.\%	& 874	\\
						& 270°C		& 0.12~at.\%	& 265 	\\
						& 330°C		& 0.12~at.\%	& 60	\\
		Zakharov \cite{ZAK97}	 	& 250°C		& 0.24~at.\%	& 942	\\
						& 300°C		& 0.24~at.\%	& 243	\\
						& 350°C		& 0.24~at.\%	& 77.7	\\
						& 400°C		& 0.24~at.\%	& 29.5	\\
		Watanabe \etal \cite{WAT04}	& 400°C		& 0.17~at.\%	& 20.9	\\
						& 450°C		& 0.17~at.\%	& 9.05	\\
		Jo and Fujikawa \cite{JO93}	& 260°C		& 0.09~at.\%	& 264	\\
						& 260°C		& 0.15~at.\%	& 440	\\
						& 300°C		& 0.09~at.\%	& 91	\\
						& 300°C		& 0.15~at.\%	& 152	\\
						& 370°C		& 0.09~at.\%	& 19.4	\\
						& 370°C		& 0.15~at.\%	& 32.4	\\
		\hline
	\end{tabular}
\end{table}

We now compare cluster dynamics predictions with all the different experimental resistivity
data available in the literature. All the experimental conditions corresponding 
to these data are given in Tab.~\ref{tab:experimental_data}.
When doing this comparison, we use
the critical size $\nSc=\infty$, \ie we assume that all clusters 
contribute to the electrical resistivity.

\subsection{Al - 0.12~at.\%~Sc}

\begin{figure}[!hbp]
	\begin{center}
		\includegraphics[width=0.8\linewidth]{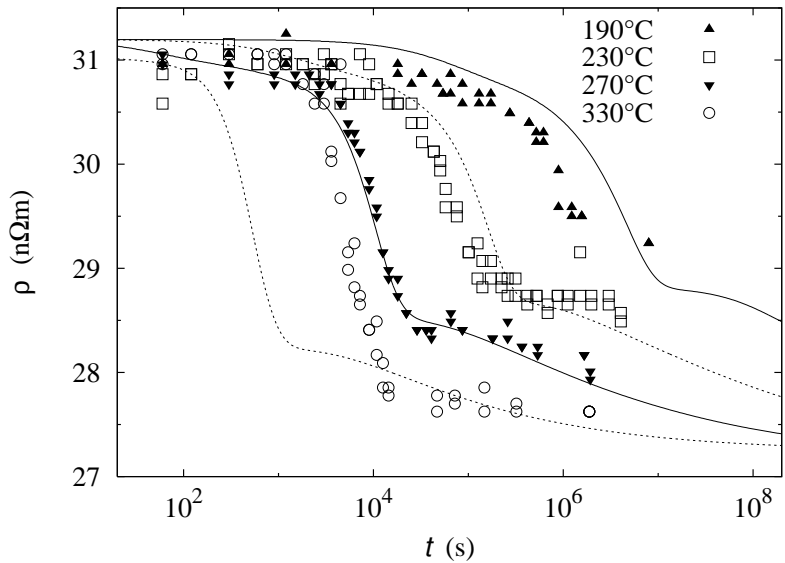}
	\end{center}
	\caption{Time evolution of the resistivity $\rho$ experimentally observed
	\cite{ROY05c} and deduced from cluster dynamics simulations for a solid solution of composition
	$x^0_{\mathrm{Sc}}=0.12$~at.\%.}
	\label{fig:Royset_resistivity}
\end{figure}

R{\o}yset and Ryum \cite{ROY05c} studied an alloy with a nominal composition\footnote{They 
actually measured by EDS analysis a slightly higher composition, $x^{0}_{\mathrm{Sc}}=0.138$~at.\%,
but cluster dynamics simulations obtained with this measured concentration do not significantly differ from 
those obtained with the nominal composition.} $x^{0}_{\mathrm{Sc}}=0.12$~at.\%.
They followed the resistivity evolution during the precipitation kinetics 
for different annealing temperatures between 190 and 470°C. 
All resistivity measurements were performed at room temperature
for which the pure Al resistivity was measured to be $\rho^{0}_{\mathrm{Al}}=27.0$~n$\Omega$m.
For temperatures lower than 330°C, they observed that precipitates remain coherent
and that precipitation is homogeneous. 
One can therefore compare their resistivity experimental data with cluster dynamics simulations
(Fig.~\ref{fig:Royset_resistivity}).
A good agreement is obtained, especially for the lowest temperatures. 
Indeed, simulations manage to reproduce the resistivity fast decrease during the nucleation 
and growth stage as well as the slower variation which follows during coarsening.
Nevertheless, for the highest temperature (330°C) which corresponds to a lower supersaturation
(Tab.~\ref{tab:experimental_data}), simulations appear to be too fast compared
to experimental data. The shape of the curve is correct, but there is a timescale shift
between the simulated and the experimental resistivity.

R{\o}yset and Ryum \cite{ROY05c} used their resistivity measurements to follow 
the Sc transformed fraction. 
In their treatment, they estimated the equilibrium solid solution from extrapolation
of resistivity measurements in the coarsening stage to infinite time. They found
that this method gave erroneous solvus estimates at low precipitation temperatures.
Our simulations (Fig.~\ref{fig:Royset_resistivity}) show that this is because 
of the excess resistivity associated with large clusters in the coarsening stage.
For high annealing temperatures (330°C for instance), the measured resistivity
is close to the equilibrium one, but for the lowest temperatures the difference cannot be neglected.
When taking into account this excess resistivity one gets  Sc transformed fractions 
different from those obtained by R{\o}yset and Ryum  (\cf appendix \ref{Royset_appendix}).
This explains why these authors deduced a wrong solubility limit for Sc in aluminum
(Fig.~11 in Ref.~\citen{ROY05c}), which they pointed out themselves, from their resistivity measurements.
As cluster dynamics parameters were obtained so as to reproduce the experimental solubility limit 
\cite{CLO05} and as our simulations manage to reproduce R{\o}yset and Ryum resistivity measurements,
we can conclude that these measurements completely agree with the assessed Sc solubility limit \cite{MUR98}.

\subsection{Al - 0.24~at.\%~Sc}

\begin{figure}[!hbtp]
	\begin{center}
		\includegraphics[width=0.8\linewidth]{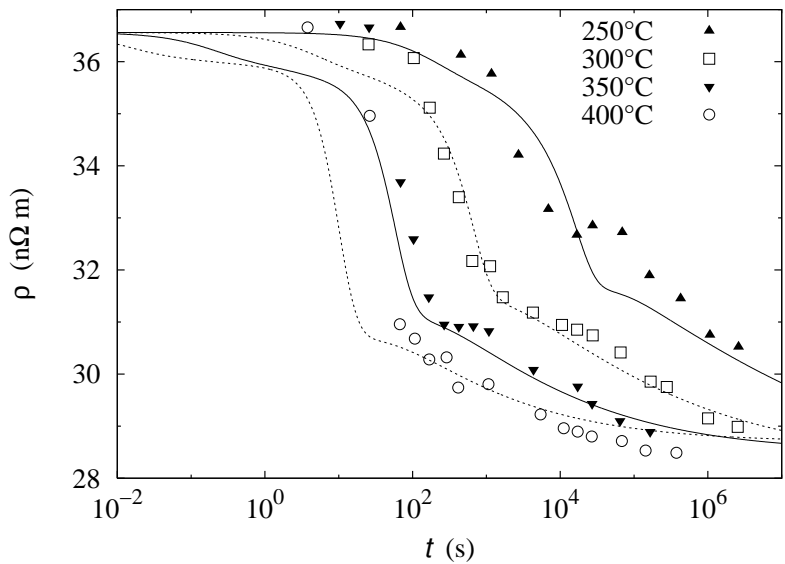}
	\end{center}
	\caption{Time evolution of the resistivity $\rho$ experimentally observed
	\cite{ZAK97} and deduced from cluster dynamics simulations for a solid solution of composition
	$x^0_{\mathrm{Sc}}=0.24$~at.\%.}
	\label{fig:Zakharov_resistivity}
\end{figure}

Zakharov \cite{ZAK97} measured the electrical resistivity variations in an alloy 
containing 0.24~at.\% Sc annealed at different temperatures between 250 and 400°C.
He did not specify at which temperature he performed these measurements, 
nor the resistivity $\rho^{0}_{\mathrm{Al}}$ of pure Al.
We assume in our simulations a resistivity $\rho^{0}_{\mathrm{Al}}=28.4$~n$\Omega$m,
corresponding therefore to a higher measurement temperature 
or a lower aluminum purity than R{\o}yset and Ryum measurements.

Cluster dynamics reproduce quite well these experimental data (Fig.~\ref{fig:Zakharov_resistivity}).
For the lowest annealing temperatures (250, 300 and 350°C), the agreement is perfect 
in all the different stages of the precipitation kinetics.
For the highest temperature (400°C), the simulated resistivity decreases 
too fast compared to the experimental one in the nucleation - growth stage. 
One should notice that the predicted evolution for this temperature is really fast
and that the resistivity drop appears at a time ($t\sim10$~s) which is too small
to be precisely observed experimentally.
Nevertheless, even for this temperature, simulations manage to reproduce
the slow decreasing of electrical resistivity during coarsening.

\subsection{Al - 0.17~at.\%~Sc}

\begin{figure}[!hbtp]
	\begin{center}
		\includegraphics[width=0.8\linewidth]{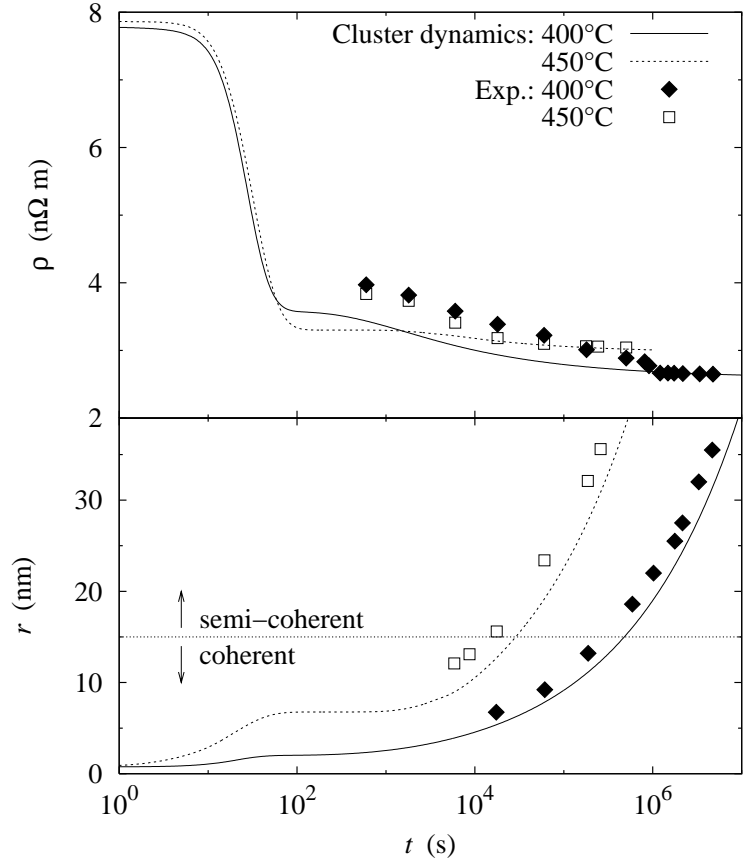}
	\end{center}
	\caption{Time evolution of the resistivity $\rho$ and of the mean precipitate radius $r$
	experimentally observed\cite{WAT04} and deduced from cluster dynamics simulations for a solid solution
	of composition 	$x^0_{\mathrm{Sc}}=0.17$~at.\%.
	The cutoff radius used to define visible precipitates in cluster dynamics is $r^{*}\sim0.75$~nm.}
	\label{fig:Watanabe_resistivite}
\end{figure}

Watanabe \etal \cite{WAT04} studied an aluminum alloy with 0.17~at.\% Sc
where they followed the resistivity 
during precipitation and measured the mean precipitate radius using transmission electron microscopy.
All measurements were performed in the coarsening stage,
after the resistivity had dropped because of precipitate nucleation and growth.
As a consequence, only a partial comparison can be made with cluster dynamics 
(Fig.~\ref{fig:Watanabe_resistivite}).
The reasonable agreement obtained between the simulated and the experimental resistivities 
is not really significant as resistivity does not vary too much in the observed
time range. Due to the high temperatures of study ($T\geq400$°C), the experimental 
as well as the simulated resistivities during coarsening are already close 
to that of the equilibrium solid solution.

Cluster dynamics manage to reproduce too the variations of the precipitate mean radius
(Fig.~\ref{fig:Watanabe_resistivite}). 
For radii greater than $\sim15$~nm, Watanabe \etal{ }observed that semi-coherent
precipitates coexist with coherent ones.
It should be pointed out that coherency loss 
is not taken into account in our simulations which handle only coherent precipitates.
Despite this assumption, cluster dynamics predictions are not too bad. 
This may indicate that the fraction of semi-coherent precipitates is small
or that these precipitates have an interface free energy not too far from the one 
of coherent precipitates.

\subsection{Al - 0.09~at.\%~Sc and Al - 0.15~at.\%~Sc}

\begin{figure}[btp]
	\begin{center}
		\subfigure[$x^0_{\mathrm{Sc}}=0.09$~at.\%]{\includegraphics[width=0.80\linewidth]{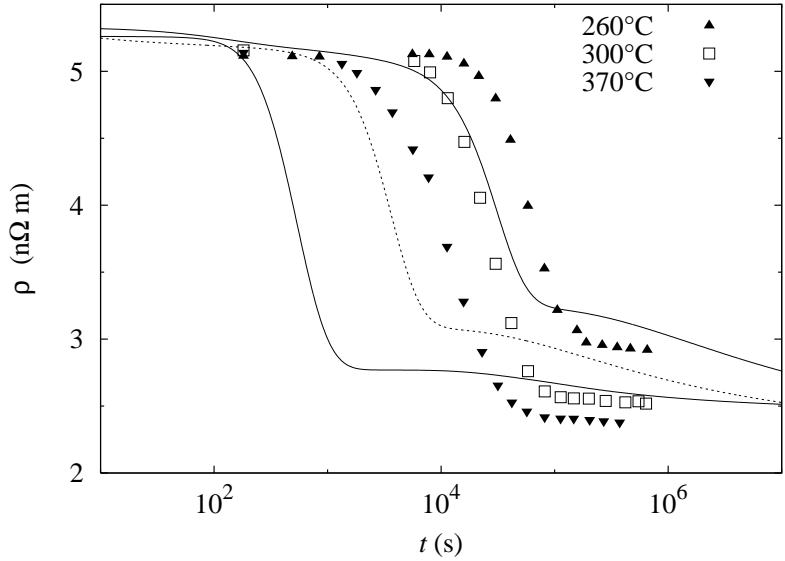}}
		\subfigure[$x^0_{\mathrm{Sc}}=0.15$~at.\%]{\includegraphics[width=0.80\linewidth]{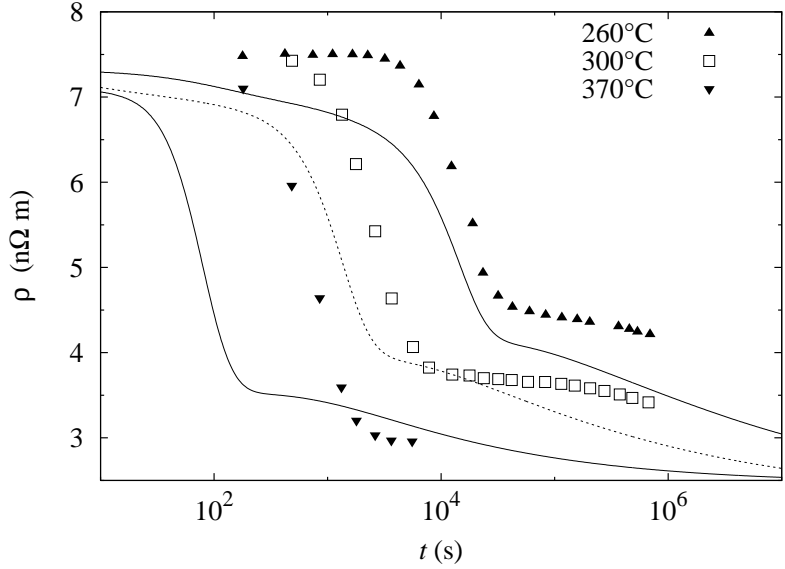}}
	\end{center}
	\caption{Time evolution of the resistivity $\rho$ experimentally observed
	\cite{JO93} and deduced from cluster dynamics simulations for two solid solutions of composition
	$x^0_{\mathrm{Sc}}=0.09$ and 0.15~at.\%.}
	\label{fig:Jo_resistivite}
\end{figure}

Jo and Fujikawa \cite{JO93} followed the resistivity variations in two different 
aluminum solid solutions containing 0.09 and 0.15~at.\% Sc.
The simulated evolutions of the electrical resistivity do not reproduce well 
their measurements (Fig.~\ref{fig:Jo_resistivite}).
The resistivity drop associated with the nucleation and growth stage 
appears to occur too fast in simulations. 
Moreover, the slow decrease of the excess resistivity during coarsening 
is not as well reproduced as with other experimental data.
Here, only a semi-quantitative agreement could be obtained.

\subsection{Summary}

Some more experimental data on resistivity measurements in Al-Sc alloys are available
in the literature, but they cannot be used for a comparison with our simulations.
Indeed, Nakayama \etal \cite{NAK97} studied an alloy containing 0.138~at.\% Sc aged
at 250, 300 and 350°C. But they normalized their measurements and did not specify
the measuring temperature nor the corresponding pure Al resistivity, thus forbidding
any use of their data.
Drits \etal \cite{DRI84} studied an alloy the Sc composition of which is larger than the Sc solubility limit
and even the eutectic composition. Some primary Al$_3$Sc precipitates may have 
appeared during the solidification and the alloy could not have been homogenized.
Therefore, one can doubt that this alloy initial state corresponds to a homogeneous supersaturated 
solid solution as assumed by our simulations.
Nevertheless, all the previously reviewed experimental data already allow 
to draw main conclusions about the ability of cluster dynamics to predict 
electrical resistivity evolution during precipitation kinetics in Al-Sc alloys.

Except for Jo and Fujikawa's data \cite{JO93} for which only a poor
agreement could be obtained, cluster dynamics manage to reproduce reasonably well 
experimental measurements of electrical resistivity \cite{ROY05c,ZAK97,WAT04}.
For the highest annealing temperatures corresponding to the lowest supersaturations,
a time shift could appear between the simulated and the experimental resistivity,
the time evolution predicted by cluster dynamics being too fast.
Nevertheless, in all cases, the global shape of the evolution is correctly predicted. 
One can guess that a better description of the resistivity drop during precipitate nucleation and growth 
may be obtained if one would calculate precisely the small cluster contributions.
But our simple model assuming a cluster contribution to the resistivity proportional to its area
already leads to good predictions. 
Most importantly, cluster dynamics manage to reproduce quantitatively 
the slow resistivity decrease during coarsening.
This decrease arises from the evolution of the large cluster distribution 
and the surface dependency assumption appears to be correct for these clusters.

An improvement of the agreement between cluster dynamics simulations and experimental data
may arise too from a more precise calculation of the emission and condensation 
rate coefficients appearing in cluster dynamics master equation as suggested
by Lépinoux\cite{LEP05,LEP06}.
But, doing so, we will lose one of the main advantages of our approach, \ie the ease 
with which the parameters of the mesoscopic modeling can be deduced 
from a very limited number of input data.
In view of its simplicity, the ability of our approach to reproduce resistivity
measurements appears more than reasonable. 

\section{Discussion}

Resistivity measurements are often combined with precipitate size determination 
using transmission microscopy so as to deduce from experimental data 
alloy parameters like the solute diffusion coefficient, the solubility limit,
and the precipitate interface free energy. This can be done with the help
of the LSW theory \cite{LIF61,WAG61}.
Indeed, Lifshitz and Slyozov\cite{LIF61} and Wagner\cite{WAG61} show that 
the precipitate mean radius varies linearly with the power $1/3$ of the time 
in the coarsening asymptotic limit.
Ardell\cite{ARD67} then extended the theory to predict the variation 
of the solid solution concentration.  
Applying his results\cite{CAL94}, the scandium concentration of the aluminum solid solution
at time $t$ should be given by
\begin{equation}
	\label{eq:coarsening_concentration}
	x_{\mathrm{Sc}}(t) = x_{\mathrm{Sc}}^{\mathrm{eq}} + \left( \kappa t \right)^{-1/3},
\end{equation}
where $x_{\mathrm{Sc}}^{\mathrm{eq}}$ is the scandium solubility in aluminum
and the rate constant $\kappa$ is\footnote{When using results obtained 
by Calderon \etal \cite{CAL94} for non pure precipitates, we assume an ideal solid solution 
and we use the fact that $x_{\mathrm{Sc}}^{\mathrm{eq}}<<x_{\mathrm{Sc}}^p$.} 
\begin{equation}
	\label{eq:coarsening_k}
	\kappa = \frac{D_{\mathrm{Sc}}}{9} 
	\left( \frac{k T}{x_{\mathrm{Sc}}^{\mathrm{eq}} \bar{\sigma} \Omega} \right)^2 x_{\mathrm{Sc}}^p .
\end{equation}
$D_{\mathrm{Sc}}$ is the scandium impurity diffusion coefficient in aluminum,
$x_{\mathrm{Sc}}^p=1/4$ is the Sc atomic fraction in the precipitate,
$\bar{\sigma}$ is the infinite radius limit of the interface free energy 
between Al$_3$Sc precipitates and aluminum,
and $\Omega=a^3/4$ is the mean atomic volume corresponding to one lattice site
($a=4.032$~{\AA} for Al).
Recently, Ardell and Ozolins \cite{ARD05} showed that the solute concentration can vary 
as the inverse square root of the time, and not the cube root like in 
Eq.~\ref{eq:coarsening_concentration}, in the case where the precipitates 
present a ragged interface. 
This is not the case in Al-Sc alloys as the Al$_3$Sc precipitate interfaces
are rather sharp as shown by our atomic simulations \cite{CLO04,CLO04T}. 
Therefore, one expects Eq.~\ref{eq:coarsening_concentration} to hold for this system.

To make use of resistivity measurements, one usually assumes that resistivity
depends linearly on the solid solution concentration. 
With the help of Eq.~\ref{eq:coarsening_concentration} and \ref{eq:coarsening_k}, 
one can then get information on the desired parameters, \ie $D_{\mathrm{Sc}}$, 
$x_{\mathrm{Sc}}^{\mathrm{eq}}$ or $\bar{\sigma}$.
We previously saw that this linear relation between electrical resistivity 
and solute concentration does not hold in a phase-separating supersaturated Al-Sc
solid solution as the resistivity has to be proportional to the cluster mean section 
(Eq.~\ref{eq:resistivity}).
It is worth looking at the error due to the fact that one identifies 
the solute concentration with the cluster mean section when exploiting 
resistivity measurements.

To do so, we need first to define the solid solution concentration 
in the cluster dynamics simulations. This is not so easy as this 
modeling technique describes the phase-separating alloy as a gas 
of solute clusters. Therefore, it does not differentiate between 
the solid solution and the precipitates at variance with precipitation 
classical descriptions like LSW theory (\cf Ref.~\citen{MAR05} for a better
understanding of the differences between cluster dynamics and classical theories).
Nevertheless, one can discriminate the solid solution and the precipitates
with the help of a threshold size $\nSc^{\mathrm{th}}$.
Below this size, clusters represents fluctuations
in the solid solution and above it they represent stable precipitates.
The solid solution concentration is thus given by
\begin{equation}
	\label{eq:concentration}
	x_{\mathrm{Sc}}(t) = \sum_{\nSc=1}^{\nSc^{\mathrm{th}}}{\nSc C_{\nSc}(t)}
\end{equation}

In a supersaturated solid solution, one natural choice for this threshold size
is the critical size $\nSc^{*}$. In cluster dynamics, this is the size
for which the condensation rate $\beta_{\nSc}$ is equal to the emission rate
$\alpha_{\nSc}$. Below this size, $\beta_{\nSc}$ is smaller than $\alpha_{\nSc}$ and
clusters have more chance to re-dissolve themselves than to grow.
This definition works as long as we do not enter in the coarsening stage.
A minimum for a size $\nSc^{\mathrm{min}}$ then appears in the cluster size distribution
($\nSc^{\mathrm{min}}=3$ for $t=10^6$ and $10^8$~s in Fig.~\ref{fig:distribution}).
Once the critical size $\nSc^{*}$ gets higher than $\nSc^{\mathrm{min}}$, 
the quantity of matter contained in clusters smaller than $\nSc^{*}$ begins
to increase artificially because of small precipitates which become unstable.
We then choose the following definition for the threshold size:
\begin{itemize}
	\item $\nSc^{\mathrm{th}}=\nSc^{*}$, as long as the cluster size 
		distribution does not show any minimum,
	\item $\nSc^{\mathrm{th}}=\min{\left(\nSc^{*},\nSc^{\mathrm{min}}\right)}$ otherwise.
\end{itemize}

\begin{figure}[!hbtp]
	\begin{center}
		\includegraphics[width=0.8\linewidth]{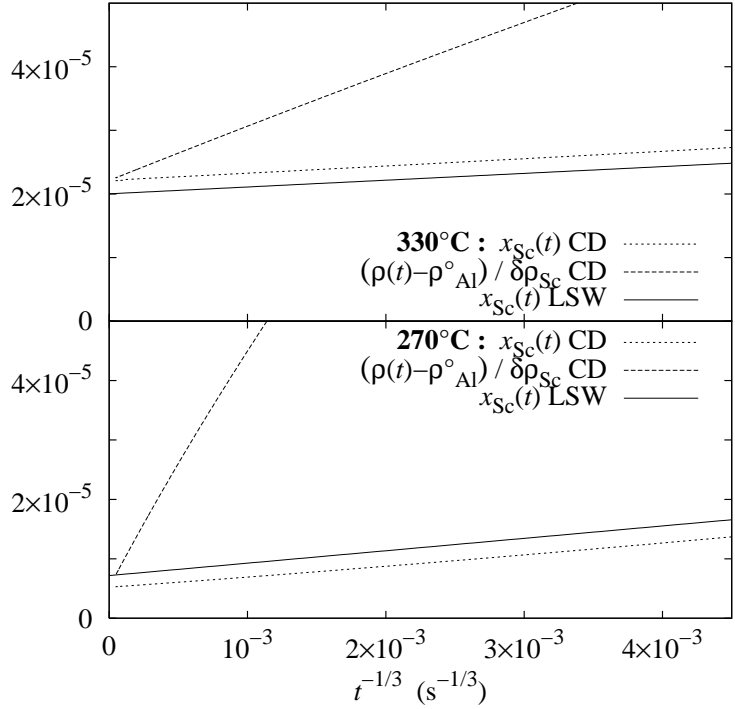}
	\end{center}
	\caption{Change as a function of $t^{-1/3}$ in the Sc concentration 
	$x_{\mathrm{Sc}}(t)$ (Eq.~\ref{eq:concentration})
	and in the normalized resistivity 
	$\left( \rho(t) - \rho^{0}_{\mathrm{Al}} \right)/\delta\rho_{\mathrm{Sc}}$ 
	(Eq.~\ref{eq:resistivity})
	simulated by cluster dynamics in an aluminum solid solution containing 0.12~at.\% Sc
	annealed at 270°C and 330°C.
	The linear relation predicted by LSW theory (Eq.~\ref{eq:coarsening_concentration})
	is shown for comparison.}
	\label{fig:lsw}
\end{figure}

Using this definition of the threshold size, we monitor the variations
of the solid solution concentration (Eq.~\ref{eq:concentration}) in clusters 
dynamics simulations for the annealing at 270 and 330°C of a solid solution containing 0.12~at.\% Sc
(Fig.~\ref{fig:lsw}). The asymptotic behavior of $x_{\mathrm{Sc}}$ 
clearly obeys a linear dependence with $t^{-1/3}$ as predicted by LSW theory.
Using the solid solubility $x^{\mathrm{eq}}_{\mathrm{Sc}}$, the diffusion coefficient
$D_{\mathrm{Sc}}$ and the interface free energy $\bar{\sigma}$ which the simulations
rely on (\cf Ref. \citen{CLO04} and \citen{CLO05} for a full description of the way 
clusters dynamics parameters were obtained), we can calculate 
coefficients entering Eq.~\ref{eq:coarsening_concentration}. 
The comparison with cluster dynamics simulations (Fig.~\ref{fig:lsw}) shows
that LSW perfectly manages to predict the asymptotic behavior of the solid 
solution concentration $x_{\mathrm{Sc}}(t)$.
On the other hand, Fig.~\ref{fig:lsw} shows that the asymptotic behavior 
of the normalized resistivity $\left( \rho(t) - \rho^{0}_{\mathrm{Al}} \right)/\delta\rho_{\mathrm{Sc}}$
differs from the one of the solid solution concentration. 
Both quantities tend to a limit close to the scandium solubility $x^{\mathrm{eq}}_{\mathrm{Sc}}$
in aluminum, but the resistivity cannot be assumed to vary linearly with $t^{-1/3}$.
If one does so, the proportionality coefficient will differ from the one predicted by
LSW theory (Eq.~\ref{eq:coarsening_k}).
As a consequence, if one assumes that the resistivity is proportional to the solid 
solution concentration and uses LSW theory to deduce alloy properties from experimental
data, the obtained solid solubility will be correct, but not the proportionality
coefficient $\kappa$. In particular, one cannot deduce from resistivity measurements
any reliable value for the precipitate interface free energy $\bar{\sigma}$ 
or the solute diffusion coefficient $D_{\mathrm{Sc}}$.
The lower the annealing temperature, the bigger the error on these parameters.
For instance, fitting with Eq.~\ref{eq:coarsening_concentration} and \ref{eq:coarsening_k}
the normalized resistivity evolution in Fig.~\ref{fig:lsw}, one would get $\bar{\sigma}=2200$ 
instead of 113~mJ.m$^{-2}$ at $T=330$°C.
This clearly illustrates that resistivity measurements cannot be used to 
determine interface free energy in a phase-separating alloy as long as 
precipitates contribute to resistivity.

Of course, this conclusion does not hold anymore when this precipitate contribution
cancels. Indeed Watanabe \etal \cite{WAT04} manages to observe a linear variation
of the resistivity with $t^{-1/3}$. This behavior was obtained once the precipitates
became incoherent. Therefore, one can reasonably assume that incoherent Al$_3$Sc 
precipitates do not contribute anymore to resistivity. 
The interface free energies they deduced from their
resistivity measurements (230~mJ.m$^{-2}$ between 400 and 450°C)
is higher than the one used in our simulations ($119\geq\bar{\sigma}\geq105$~mJ.m$^{-2}$
between 200 and 500°C) which was deduced from the atomic diffusion model
of Ref. \citen{CLO04,CLO06}. This is in agreement with the fact that incoherent
precipitates should have a higher interface free energy than coherent ones.

\section{Conclusions}

Cluster dynamics has been shown to be able to reproduce
electrical resistivity variations during precipitation in Al-Sc alloys.
One should recall that the only input parameters required by
this modeling technique are the solute diffusion coefficients,
the precipitate/aluminum interface free energy and the resistivity
increase per solute atom.
None of these parameters was adjusted to fit the experimental data.
Indeed, the experimental data measured by Fujikawa \etal \cite{JO93,FUJ79} 
was used for the resistivity increase with solute content ($\delta\rho_{\mathrm{Sc}}=3400$~n$\Omega$m).
Good predictions were obtained with this experimental value.
As for the solute diffusion coefficients and the interface free energies,
they were directly deduced\cite{CLO05} from an atomic model previously
developed for Al-Zr-Sc alloys\cite{CLO04,CLO06}.
The good agreement obtained between our resistivity simulations and experimental
data allows then us to stress the correctness of the Al$_3$Sc interface free energies 
used in our simulations.
For temperatures ranging between 200 and 500°C, this interface free energy 
corresponding to coherent Al$_3$Sc precipitates is varying between 119 and 105~mJ.m$^{-2}$.

One of the key assumptions of our simulations is that all clusters 
contribute to electrical resistivity and that each cluster contribution
is proportional to its area. 
Although this assumption may look crude, it leads to quantitative predictions.
In particular, it manages to reproduce the resistivity excess and its slow decrease 
during coarsening. 
This excess resistivity mainly arises from large clusters contributions
whereas the solid solution contribution can be neglected.
As a consequence, resistivity measurements during coarsening do not really allow
to follow the solid solution concentration.
In particular, resistivity do not obey LSW theory in a phase-separating system
like supersaturated Al-Sc alloys at variance with the solid solution concentration.
This involves that one cannot deduce from these measurements
correct values of the precipitate interface free energy 
or of the solute diffusion coefficients.
On the other hand, solubility limits obtained from resistivity experiments 
are correct as both the normalized resistivity and the solid solution concentration tend
to the same value for long enough annealing times.

\appendix
\section{Reinterpretation of R{\o}yset experimental data}
\label{Royset_appendix}

\begin{figure}[!hbtp]
	\begin{center}
		\includegraphics[width=0.49\linewidth]{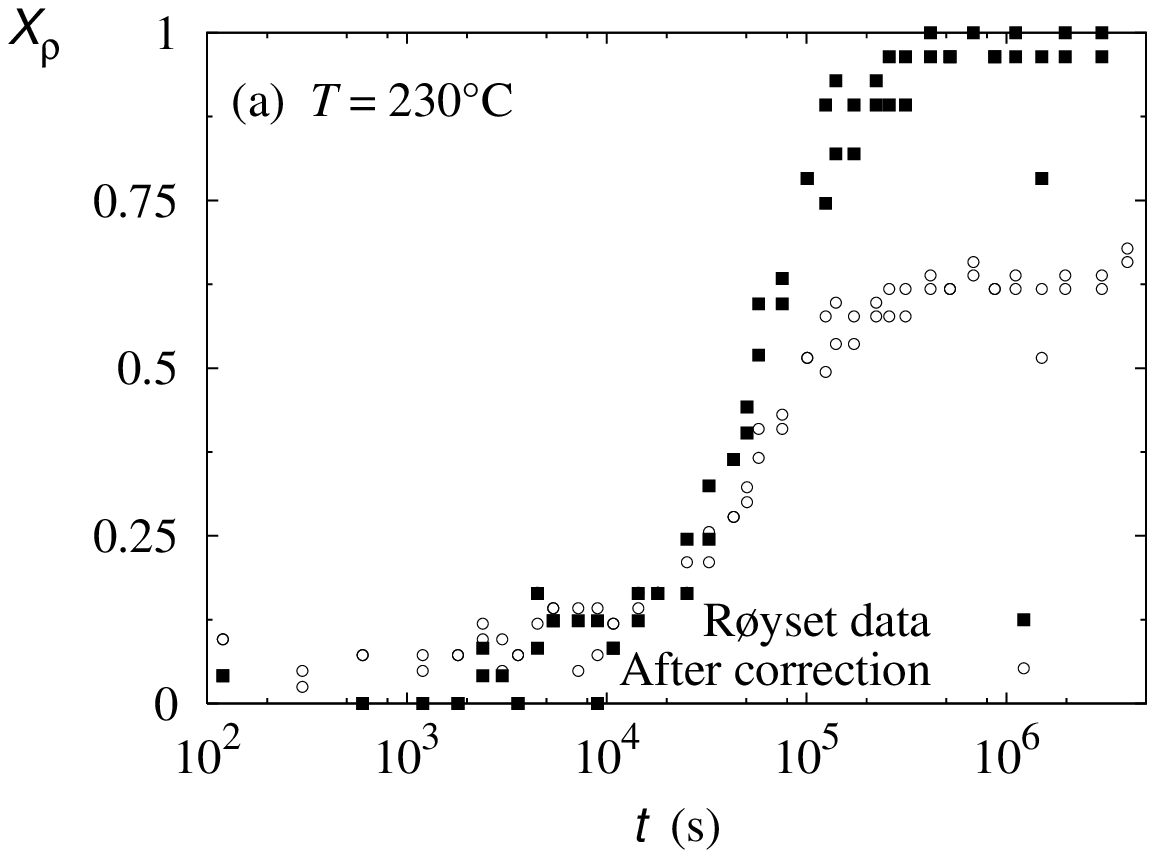}
		\hfill
		\includegraphics[width=0.49\linewidth]{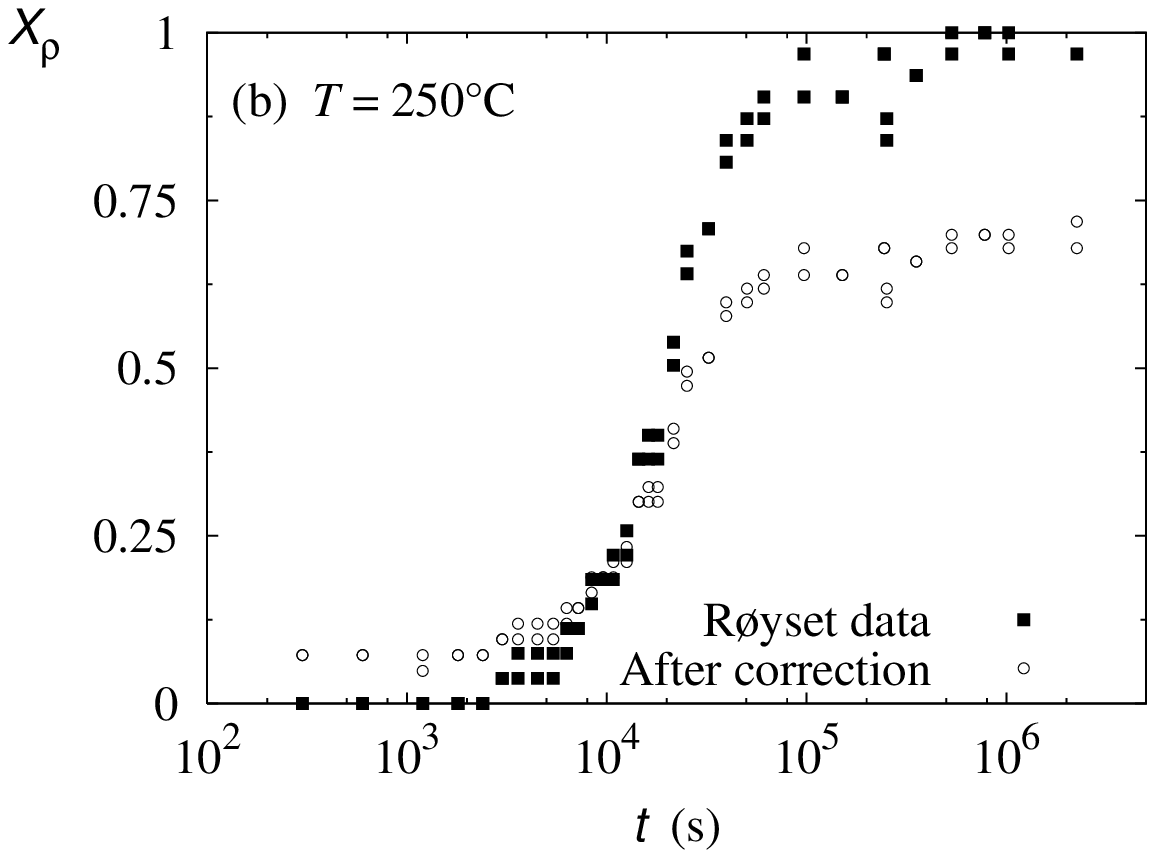}
		\includegraphics[width=0.49\linewidth]{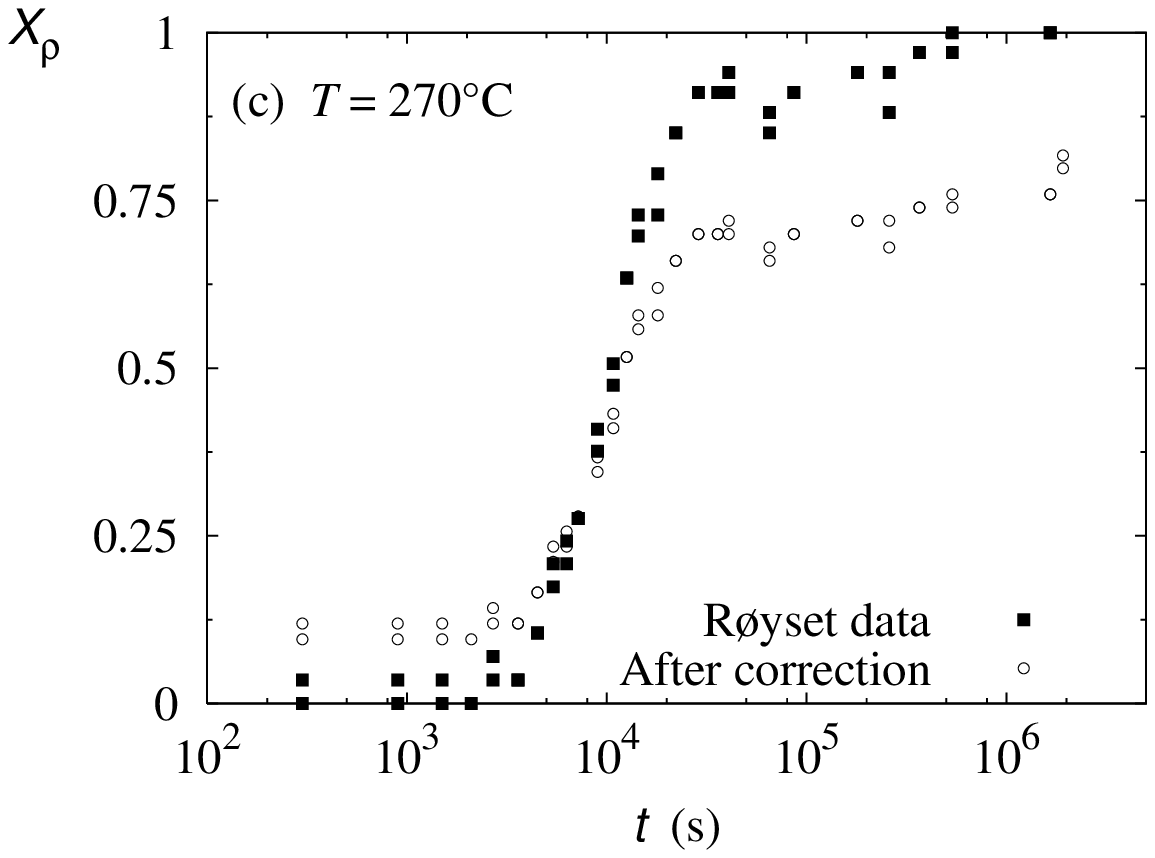}
		\hfill
		\includegraphics[width=0.49\linewidth]{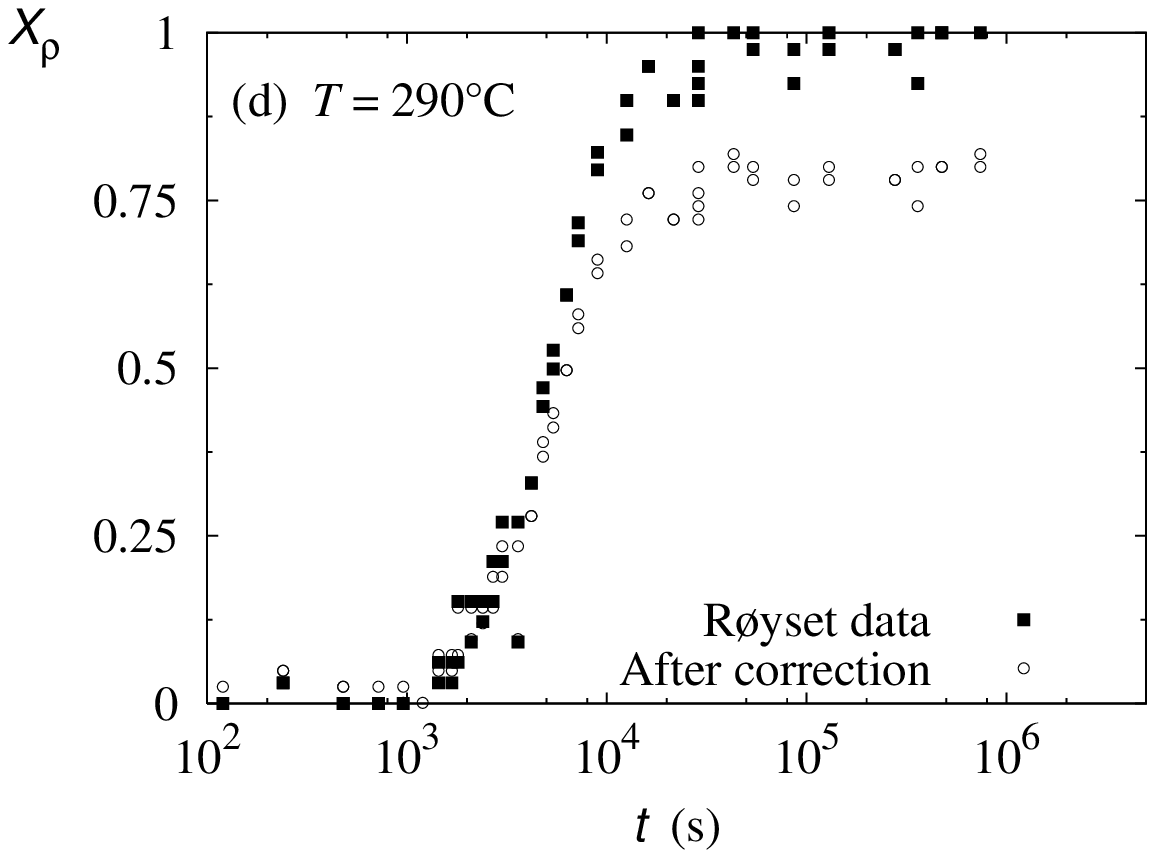}
		\includegraphics[width=0.49\linewidth]{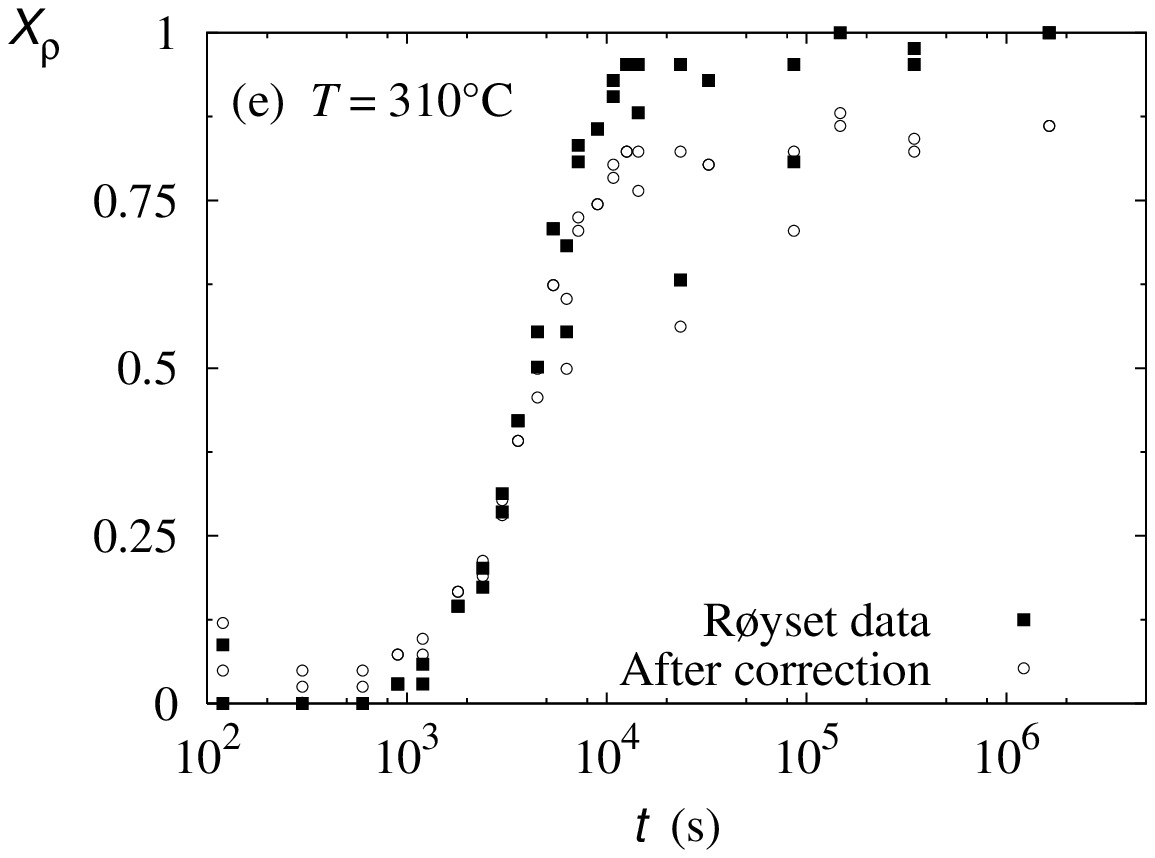}
		\hfill
		\includegraphics[width=0.49\linewidth]{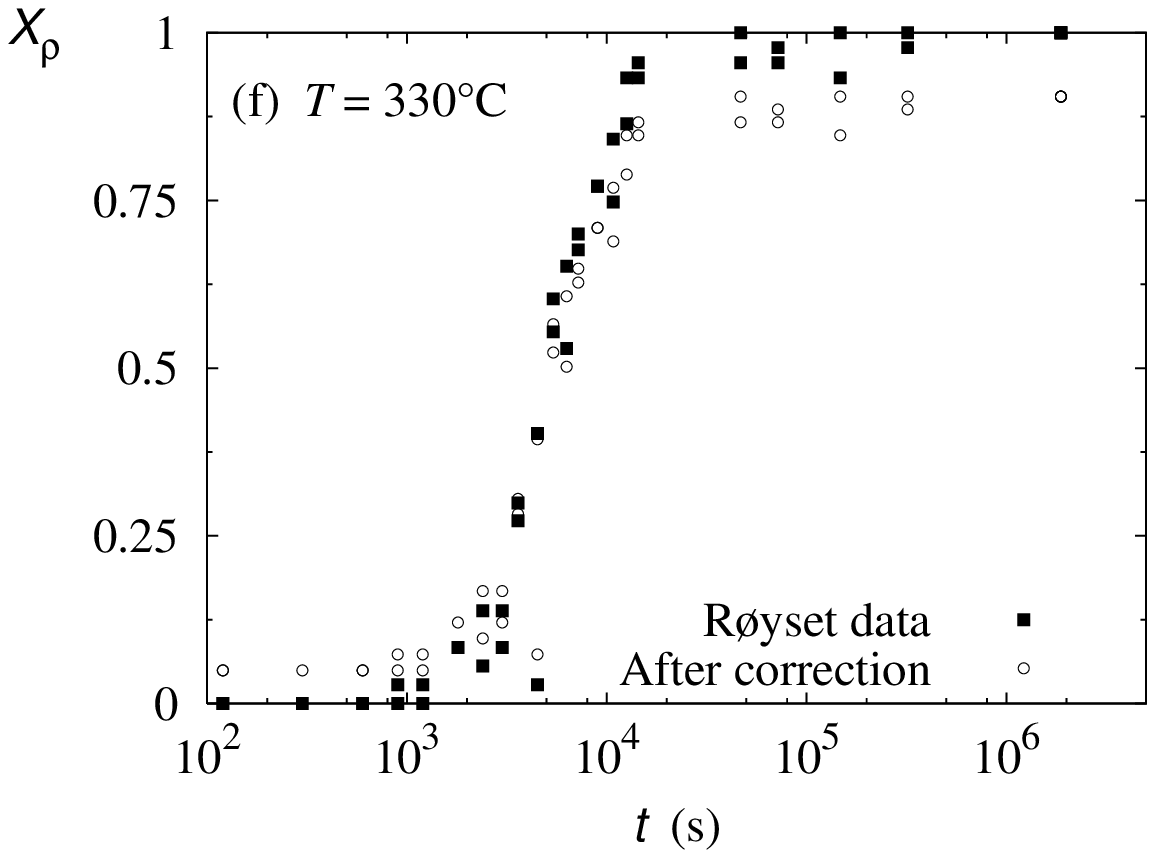}
	\end{center}
	\caption{Evolution with time of the precipitated fraction $X_{\rho}$
	in an aluminum solid solution  of composition $x^0_{\mathrm{Sc}}=0.12$~at.\%.
	$X_{\rho}$ has been deduced from experimental resistivity measurements\cite{ROY05c}
	considering the equilibrium resistivity is reached (R{\o}yset data)
	or not (after correction) for the longest annealing time.}
	\label{fig:Royset_avancement}
\end{figure}

Usually, one uses electrical resistivity measurements to define the 
fraction of precipitated solute
\begin{equation}
	X_{\rho}(t) = \frac{\rho(t)-\rho^0}{\rho^{\mathrm{eq}}-\rho^0},
	\label{eq:avancement}
\end{equation}
where $\rho(t)$ is the resistivity of the solid solution measured at time $t$
and $\rho^0$ and $\rho^{\mathrm{eq}}$ are respectively the initial and equilibrium resistivities.
If the resistivity was truly proportional to the solute concentration 
like it can be assumed in an under-saturated solid solution, 
this definition would be equivalent to the one obtained from considering the solid solution concentration
instead of the resistivity. But in a phase-separating system like supersaturated
Al-Sc solid solution, both definitions lead to different quantities.

In their work \cite{ROY05c}, R{\o}yset and Ryum used Eq.~\ref{eq:avancement}
to define the precipitated fraction.
Assuming that resistivity varies linearly with $t^{-1/3}$ in the coarsening stage,
extrapolation of data to infinite time was used
to estimate the equilibrium resistivity $\rho^{\mathrm{eq}}$.
This leads to a correct value of $\rho^{\mathrm{eq}}$ 
for the higher annealing temperatures but not for the lower ones ($T\leq330^{\circ}$C).
Indeed, as we previously saw, one cannot assume that resistivity follows
the same time dependence as the solid solution concentration due to 
the large cluster contributions to electrical resistivity.
Therefore $\rho^{\mathrm{eq}}$ cannot be obtained from such an extrapolation 
to infinite time, and it has to be calculated 
from the equilibrium Sc solubility in aluminum 
and the linear relation\cite{FUJ79,JO93} between resistivity 
and concentration existing in dilute solid solutions leading to
\begin{equation}
	\rho^{\mathrm{eq}} = \rho^0_{\mathrm{Al}} \
		+ \delta\rho_{\mathrm{Sc}} x_{\mathrm{Sc}}^{\mathrm{eq}},
	\label{eq:resistivity_appendix}
\end{equation}
The precipitated fraction $X_{\rho}$ obtained calculating $\rho^{\mathrm{eq}}$ in this way 
differs from the one obtained by R{\o}yset and Ryum
who assumed that $\rho^{\mathrm{eq}}$ corresponds to the resistivity
measurements for the longest annealing time (Fig.~\ref{fig:Royset_avancement}).
This is a clear manifestation of the large cluster contributions
to electrical resistivity.

\begin{ack}
	The authors are grateful to Dr. R{\o}yset and Dr. Watanabe 
	for providing experimental data.
	They want to thank too Dr. Martin and Dr. Soisson for their
	careful reading of the manuscript,
	as well as Dr. Limoge, Dr. Marinica and Dr. Proville 
	for useful discussions on electron scattering.
\end{ack}

\bibliographystyle{elsart-num}
\bibliography{./clouet2007}

\end{document}